\documentclass[preprintnumbers,article,amsmath,amssymb,floatfix,10pt,prd,twocolumn,
superscriptaddress,nofootinbib]{revtex4-2}
\usepackage{bm}
\usepackage{amsfonts}
\usepackage{latexsym}
\usepackage[latin1]{inputenc}
\usepackage{graphicx}
\usepackage{amsmath}
\usepackage{palatino}
\usepackage{mathpazo}
\usepackage{textcomp}
\usepackage{float}
\usepackage{booktabs}
\usepackage{dcolumn}
\usepackage{ragged2e}
\usepackage{hyperref}
\hypersetup{colorlinks,citecolor=blue}
\hypersetup{colorlinks=true,linkcolor=red,filecolor=magenta,    urlcolor=blue}
\usepackage{amsmath}
\usepackage{xcolor}
\usepackage{orcidlink}
\usepackage{epsfig}
\usepackage{caption}
\usepackage{subcaption}
\usepackage{commath}
\def\jnl@style{\it}
\def\aaref@jnl#1{{\jnl@style#1}}

\def\aaref@jnl#1{{\jnl@style#1}}

\def\aj{\aaref@jnl{AJ}}                   
\def\apj{\aaref@jnl{ApJ}}                 
\def\apjl{\aaref@jnl{ApJ}}                
\def\apjs{\aaref@jnl{ApJS}}               
\def\apss{\aaref@jnl{Ap\&SS}}             
\def\aap{\aaref@jnl{A\&A}}                
\def\aapr{\aaref@jnl{A\&A~Rev.}}          
\def\aaps{\aaref@jnl{A\&AS}}              
\def\mnras{\aaref@jnl{Mon.~Not.~Roy.~Astron.~Soc.}}             
\def\prd{\aaref@jnl{Phys.~Rev.~D}}        
\def\prc{\aaref@jnl{Phys.~Rev.~C}}  
\def\prl{\aaref@jnl{Phys.~Rev.~Lett.}}    
\def\qjras{\aaref@jnl{QJRAS}}             
\def\skytel{\aaref@jnl{S\&T}}             
\def\ssr{\aaref@jnl{Space~Sci.~Rev.}}     
\def\zap{\aaref@jnl{ZAp}}                 
\def\nat{\aaref@jnl{Nature}}              
\def\aplett{\aaref@jnl{Astrophys.~Lett.}} 
\def\apspr{\aaref@jnl{Astrophys.~Space~Phys.~Res.}} 
\def\physrep{\aaref@jnl{Phys.~Rep.}}      
\def\physscr{\aaref@jnl{Phys.~Scr}}       
\def\commat{\aaref@jnl{Comm.~Math.~Phys.}}              
\def\science{\aaref@jnl{Science}}               
\def\cqg{\aaref@jnl{Classical Quant.~Grav.}}            
\def\jpcs{\aaref@jnl{JPCS}}                                     
\def\ijmpd{\aaref@jnl{Int.~J.~Mod.~Phys.~D}}                    
\def\grg{\aaref@jnl{Gen.~Relat.~Gravit.}}               
\def\rpp{\aaref@jnl{Rep.~Prog.~Phys.}}          
\def\npa{\aaref@jnl{Nucl.~Phys.~A}}        
\def\lrr{\aaref@jnl{Living Rev.~Rel.}}                   
\def\jcap{\aaref@jnl{J.~Cosmology Astropart.~Phys.}}    
\def\rmp{\aaref@jnl{Rev.~Mod.~Phys.}}   
\def\epjc{\aaref@jnl{Eur.~Phys.~J.~C}} 
\def\plb{\aaref@jnl{~Phy.~Lett.~B}} 
\def\mpla{\aaref@jnl{Mod.~Phy.~Lett.~A}} 
\def\arxiv{\aaref@jnl{arxiv.org}}

\allowdisplaybreaks[1]

\begin{document}
\color{black}       
\title{Generalized Chaplygin gas and accelerating universe in $f(Q,T)$ gravity}

\author{Gaurav N. Gadbail \orcidlink{0000-0003-0684-9702}}
\email{gauravgadbail6@gmail.com}
\affiliation{Department of Mathematics, Birla Institute of Technology and
Science-Pilani,\\ Hyderabad Campus, Hyderabad-500078, India.}

\author{Simran Arora\orcidlink{0000-0003-0326-8945}}
\email{dawrasimran27@gmail.com}
\affiliation{Department of Mathematics, Birla Institute of Technology and
Science-Pilani,\\ Hyderabad Campus, Hyderabad-500078, India.}

\author{P.K. Sahoo\orcidlink{0000-0003-2130-8832}}
\email{pksahoo@hyderabad.bits-pilani.ac.in}
\affiliation{Department of Mathematics, Birla Institute of Technology and
Science-Pilani,\\ Hyderabad Campus, Hyderabad-500078, India.}
%

\begin{abstract}
The generalized Chaplygin gas (GCG), which has an unusual perfect fluid equation of state, is another promising candidate for dark energy. We investigate the GCG scenario coupled with a baryonic matter in a newly suggested $f(Q,T)$ gravity, an arbitrary function of non-metricity $Q$ and the trace of energy-momentum tensor $T$. We consider the functional form of $f(Q, T)$ as a linear combination of $Q$ and an arbitrary function of $T$, denoted by $h(T)$. Furthermore, we obtain two  different functional forms of the $f(Q,T)$ model under high pressure and high-density scenarios of GCG.  We also test each model with the recent Pantheon supernovae data set of 1048 data points, Hubble data set of 31 points, and baryon acoustic oscillations. The deceleration parameter is constructed using $OHD+SNeIa+BAO$, predicting a transition from decelerated to accelerated phases of the universe expansion. Also, the equation of state parameter acquires a negative behavior depicting acceleration. Finally, we analyze the statefinder diagnostic to discriminate between the GCG and other dark energy models.

\end{abstract}

\maketitle

\date{\today}

\section{Introduction}
The accelerated expansion of the universe has yet to be assigned to a universally acknowledged form of dark energy, responsible for the alleged acceleration, despite being known for more than two decades. As a result, the search for dark energy has continued in all its forms. The standard model of cosmology, named as the $\Lambda$CDM model, is supported by a number of observations \cite{Perlmutter/1999,Riess/1998,Riess/2004,Spergel/2007,Koivisto/2006,Daniel/2008}. The cosmological constant having a negative constant equation of state, provides the simplest understanding of dark energy, which is responsible for the late-time acceleration in the framework of general relativity (GR). Despite its simplicity and flexibility to work with a wide range of up-to-date observational data, the $\Lambda$CDM paradigm still has theoretical and observational issues that cannot be solved \cite{Weinberg/1989,Steinhardt/1999}. The well-known cosmological constant concerns are among the theoretical challenges. Alternative possibilities for DE has been offered in the literature, such as quintessence \cite{Ratra/1988,Caldwell/1998}, phantom \cite{Caldwell/2002,Caldwell/2003}, quintom \cite{Feng/2005,Guo/2005}, and decaying vacuum models \cite{Xu/2011,Tong/2011}, etc.\\
With the purpose of solving the DE problem, several authors \cite{Kamenshchik/2001,Gorini/2004,Salahedin/2020,Ferreira/2018} introduced the Chaplygin gas (CG), which has an unusual perfect-fluid equation of state, $p=-\frac{A}{\rho}$, where $p$ and $\rho$ are the pressure and energy density, respectively, and $A$ is a positive constant. The model could give a unified description of dark matter and dark energy. The most unexpected feature of this model is that, CG acts as a pressureless dark matter in the early universe, while late time acts as a cosmological constant. There are problems with the CG model, though. It cannot describe how structures are made, and it has problems with the cosmological power spectrum \cite{Bean/2003,Sandvik/2004}.\\
To address this problem, Bento et al., \cite{Bento/2002} generalized the Chaplygin gas into the generalized Chaplygin gas (GCG).  The GCG is defined by $p=-\frac{A}{\rho^{\alpha}}$ with $0<\alpha \leqslant1$ and $A>0$. The case with  $\alpha=0$ and finite $A$, on the other, is identical to the $\Lambda$CDM model. It has been demonstrated that after phantom-like dark energy is ruled out, this equation of state uniquely characterizes an interacting mixture of decaying dark matter and a  cosmological constant. In this case, the difficulty of preserving unphysical features in the cosmic power spectrum can be avoided \cite{Bento/2004}. As a result, the model explains the cosmic expansion history of the universe that transitioned from decelerating to an accelerating phase \cite{Barreiro/2008}. Many authors have looked at the GCG model in the literature and found it to be accurate. They demonstrated that the GCG model may be viewed as an interactive variant of $\Lambda$CDM \cite{Barreiro/2008,Zhang/2006,Salahedin/2020,Ferreira/2018,Dindam/2014,Ebadi/2016} .\\
Another proposal to understand the dark energy problem is to modify the gravitational Einstein-Hilbert action known as modified gravity theories, including $f(R)$ theory \cite{Buchdahl/1970,Starobinsky/2007}, $f(T)$ theory \cite{Capozziello/2011,Cai/2016}, $f(R,T)$ theory \cite{Harko/2011,Moraes/2017}. Theories of gravity can be classified into three categories based on their connection. They are as follows: the Levi-Civita connection and curvature; the metric tetrads and torsion; and the third employs a torsion and curvature free symmetric teleparallel connection that is not metric compatible. By requiring that the curvature vanishes, and the connection is torsionless, the remaining gravitational interaction is endowed by non-metricity $Q$. Here, $Q$ geometrically describes the variation of the length of a vector in the parallel transport. In the present article, we will work with the extension of the recently introduced $f(Q)$ theory of gravity \cite{Jimenez/2018,Harko/2018}, known as $f(Q,T)$ gravity \cite{Xu/2019}, where $Q$ is the non-metricity and $T$ is the trace of the energy-momentum tensor. Despite its newness, the $f(Q,T)$ theory has some intriguing and useful applications in the literature. The first cosmological implications in $f(Q,T)$ gravity appear in Reference \cite{Xu/2019}, while late-time accelerated expansion with observational constraints can be seen in \cite{Arora/2020,Arora/2021}. To get in touch with other works in $f(Q,T)$ theory, one can check references \cite{Bhattacharjee/2020,Najera/2022,Arora/2022,Najera/2021}.

Inspired by the newly proposed theory, the main purpose of this study is to investigate the generalized Chaplygin gas conjunction with a baryonic matter in the $f(Q,T)$ gravity theory. To achieve plausible precise solutions, we will start with an equation of state for the GCG and pressureless baryonic matter, which reduces to the $\Lambda$CDM scenario in the approximate cosmological limits. Anyhow, one can check \cite{Elmardi/2016} the Chaplygin gas solutions in the $f(R)$ gravity, which are generally quadratic in $R$ having adequate $\Lambda$CDM solutions as limiting cases;  the cosmological behavior of the GCG in $f(R, T)$ gravity endowed with isotropic and homogeneous FLRW space-time \cite{Shabani/2017}.  The GCG interacting with $f(R, T)$ gravity in the presence of shear and bulk viscosities has also been studied in \cite{Baffou/2017}.\\
The outline is as follows: In section \ref{section 2}, we introduce the $f(Q, T)$ gravity formalism and their corresponding field equations. In section \ref{section 3}, we describe the observational data and constrain the model parameters using $OHD+SNeIa+BAO$. We analyze the behavior of different cosmological parameters such as deceleration parameter and equation of state in section \ref{section 4}. In section \ref{section 5}, we observed the behavior of statefinder parameters on the values constrained by the observational data to differentiate between dark
energy models. Lastly, we discuss our results in section \ref{section 6}
     
\section{The $f(Q,T)$ Theory}
\label{section 2}
In this section, we briefly present the $f(Q,T)$ gravity in the presence of pressureless baryonic matter 
and generalized Chaplygin gas (GCG) as the matter contents. We start with the $f(Q,T)$ gravity, where the action is given by \cite{Xu/2019}
   
\begin{equation}
\label{1}
S=\int \sqrt{-g}\left(\frac{1}{16\pi G}f(Q,T^{(b,G)})+\mathcal{L}_m^{(b,G)}\right) d^4x,
\end{equation}
where, $Q$ is the non-metricity and $T^{(b,G)}=g^{\alpha \beta}\left(T^{(b)}_{\alpha\beta}+T^{(G)}_{\alpha\beta}\right)$ is the trace of the energy momentum tensor of the baryonic matter (superscript $b$) and GCG (superscript $G$). The $\mathcal{L}_m^{(b,G)}= \mathcal{L}_m^{(b)}+\mathcal{L}_m^{(G)}$ is the total matter Lagrangian. The scalar non-metricity $Q$ is defined as
  
\begin{equation}
\label{2}
Q\equiv- g^{\mu\nu}\left(L^\lambda_{\,\,\, \beta\mu}L^\beta_{\,\,\, \nu\lambda}-L^\lambda_{\,\,\, \beta\lambda}L^\beta_{\,\,\, \mu\nu}\right),
\end{equation}
where $L^\lambda_{\,\,\,\beta\mu}$ is the deformation tensor
\begin{equation}
\label{3}
L^\lambda_{\,\,\, \beta\mu}=-\frac{1}{2}g^{\lambda\gamma}\left(\nabla_{\mu}g_{\beta\gamma}+\nabla_{\beta}g_{\gamma\mu}-\nabla_{\gamma}g_{\beta\mu}\right).
\end{equation}

Introducing the superpotential $P_{\,\,\,\,\mu\nu}^{\lambda}$,
\begin{equation}
\label{4}
P_{\,\,\,\,\mu\nu}^{\lambda}=-\frac{1}{2}L^\lambda_{\,\,\,\mu\nu}+\frac{1}{4}\left(Q^{\lambda}-\tilde{Q^{\lambda}}\right)g_{\mu\nu}-\frac{1}{4} \delta^{\lambda}_{\,\,(\mu}Q_{\nu)},
\end{equation} 
giving the scalar non-metricity $Q=-Q_{\lambda\mu\nu}P^{\lambda\mu\nu}$. Further, we define the trace of non-metricity tensor as $Q_{\lambda}= {Q_{\lambda}^{\,\,\,\, \mu}}_{\,\,\,\mu}$ and $\tilde{Q_{\alpha}}= Q^{\mu}_{\,\,\, \lambda \mu}$.

Varying the action \eqref{1} with respect to the metric, one obtains the following gravitational field equation of the $f(Q,T)$ gravity:
\begin{multline}
\label{5}
-\frac{2}{\sqrt{-g}}\nabla_{\lambda}\left(f_{Q}\sqrt{-g}\,P^{\lambda}_{\,\,\,\mu\nu}\right)-\frac{1}{2}f\,g_{\mu\nu}+f_{T}\left(T_{\mu\nu}+\Theta_{\mu\nu}\right)\\
-f_{Q}\left(P_{\mu\lambda\beta}Q_{\nu}^{\,\,\,\lambda\beta}-2Q^{\lambda \beta}_{\,\,\, \, \,\, \mu}P_{\lambda\beta\nu}\right)=8\pi G T_{\mu\nu}.
\end{multline}
Here, $T_{\mu \nu}$ and $\theta_{\mu\nu}$ defined as
\begin{equation}
\label{6}
T_{\mu\nu}\equiv-\frac{2}{\sqrt{-g}}\frac{\delta(\sqrt{-g}\mathcal{L}_m^{(b,G)})}{\delta g^{\mu\nu}},\,\, \theta_{\mu\nu}\equiv g^{\alpha\beta}\frac{\delta T_{\alpha\beta}}{\delta g^{\mu\nu}}
\end{equation} 
\begin{equation}
\label{7}
f_T\equiv \frac{\partial f(Q,T)}{\partial T},
f_Q\equiv\frac{\partial f(Q,T)}{\partial Q}
\end{equation}
respectively. 
Assuming a perfect fluid and spatially flat Friedmann-Lema\^{i}tre-Robertson-Walker (FLRW) metric,
\begin{equation}
\label{8}
ds^2=-dt^2+a^2(t)\left( dx^2+dy^2+dz^2\right).
\end{equation}
Using above FLRW metric with Eqs.\eqref{2} and \eqref{3}, we get the relation $Q=6H^2$, where $H=\frac{\dot{a}}{a}$ is the Hubble parameter and $a(t)$ is a cosmic scale factor. Substituting the above FLRW metric into the gravitational field equations \eqref{5}, we obtain the following:
\begin{equation}
\label{9}
3H^2=8\pi G \rho_{eff}=\frac{f}{4F}-\frac{4\pi}{F}\left((1+\tilde{G})\rho+\tilde{G}p\right),
\end{equation}
\begin{multline}
\label{10}
2\dot{H}+3H^2=-8\pi G p_{eff}=\frac{f}{4F}-\frac{2\dot{F}H}{F}\\
+\frac{4\pi}{F}\left((1+\tilde{G})\rho+(2+\tilde{G})p\right),
\end{multline}
where $8\pi\tilde{G}=f_T$ and $F=f_Q$. Eqs. \eqref{9} and \eqref{10} shows that the effective thermodynamical quantities satisfy the conservation equation \cite{Xu/2019}. \\
Now, we start with a generic functional form of $f(Q,T)$ read as \cite{Shabani/2017, Shabani/2013}
\begin{equation}
\label{11}
f(Q,T)=Q+h(T^{(b,G)})
\end{equation}   
where $h(T^{(b,G)})=h_1(T^{(b)})+h_2(T^{(G)})$. Depending on the matter content, the considered functional form can be viewed as a simple modification to GR. Furthermore, rather than using sophisticated numerical approaches, this model produces exact solutions.\\
For the choice of above functional form and using $Q=6H^{2}$, we get Eq.\eqref{9} as follows:
\begin{equation}
\label{12}
-3H^2=-\frac{1}{2}h(T)+(8\pi G +h'(T))T_{\mu\nu}-  h'(T)pg_{\mu\nu},
\end{equation}
and we obtained
\begin{equation}
\label{13}
3H^2=\sum_{i=1}^2\left(\frac{1}{2}h_i-8\pi G\rho_i-h'_i(\rho_i+p_i)\right).
\end{equation}
Varying summation in the above equation, we get
\begin{multline*}
3H^2=\frac{1}{2}h_1(T^{(b)})+\frac{1}{2}h_2(T^{(G)})-8\pi G(\rho^{(b)}+\rho^{(G)})\\
-h'_1(T^{(b)})(\rho^{(b)}+p^{(b)})-h'_2(T^{(G)})(\rho^{(G)}+p^{(G)})
\end{multline*}

Concerning the Bianchi identity, the effective energy-momentum tensor is not conserved in $f(Q, T)$ gravity. Thus, by using the conservation of the energy-momentum tensor of all matter and knowing that $\nabla^{\mu}T^i_{\mu\nu}=0$ (where i=b, G), the Eq.\eqref{12} leads to the following constraint equation for a pressureless baryonic matter:
\begin{equation}
\label{14}
\dot{h}'_1\rho^{(b)}+\frac{1}{2}h'_1\dot{\rho}^{(b)}=0,
\end{equation}
where $T^{(b)}=-\rho^{(b)}$. Solving Eq.\eqref{14} gives the following:
\begin{equation}
\label{15}
h_1(T^{(b)})=c_1^{(b)}\sqrt{-T^{(b)}}+c_2^{(b)},
\end{equation}
where $c_1^{(b)}$ and $c_2^{(b)}$ are integration constants.\\
The equation of state relates the GCG background \cite{Lu/2009,Ferreira/2018,Salahedin/2020} fluid with its energy density and pressure as
\begin{equation}
\label{16}
p^{(G)}=-\frac{A}{{\rho^{(G)}}^{\alpha}},
\end{equation}
where $A>0$ and $0< \alpha \leq 1$ are free parameters. When $\alpha = 1$, it reduces to the CG scenario. Combining the energy conservation equation for generalised chaplygin gas $\dot{\rho}^{(G)}+3H(\rho^{(G)}+p^{(G)})=0$ and Eq.\eqref{16}, the energy density of the GCG fluid can be expressed as
\begin{equation}
\label{17}
\rho^{(G)}=\rho_0\left(k+\left(1-k\right)a^{-3(1+\alpha)}\right)^{\frac{1}{\alpha+1}}
\end{equation}
where $k=\frac{A}{\rho_0^{\alpha+1}}>0$.\\
We discover the solution of the constraint equation for GCG in two scenarios. First, GCG in the high pressure regimes i.e.$p^{(G)}>>\rho^{(G)}$ and second, GCG in the high density regimes i.e. $\rho^{(G)}>>p^{(G)}$. In these cases, we will consider $\alpha$ as arbitrary. The following solutions are
\begin{equation}
\label{18}
h_2(T^{(G)})=\frac{2}{3}c_1^{(G)}(-T^{(G)})^{\frac{3}{2}}+c_2^{(G)},
\end{equation}
\begin{equation}
\label{19}
h_2(T^{(G)})=2c_1^{(G)}(-T^{(G)})^{\frac{1}{2}}+c_2^{(G)}.
\end{equation}
respectively.
We obtained two functional forms of $f(Q,T)$ as
\begin{equation}
\label{20}
f(Q,T)=Q+c_1^{(b)}\sqrt{-T^{(b)}}+\frac{2}{3}c_1^{(G)}(-T^{(G)})^{\frac{3}{2}}+\Lambda^{(b,G)}.
\end{equation}
for $p^{(G)}>>\rho^{(G)}$.
\begin{equation}
\label{21}
f(Q,T)=Q+c_1^{(b)}\sqrt{-T^{(b)}}+2c_1^{(G)}(-T^{(G)})^{\frac{1}{2}}+\Lambda^{(b,G)}.
\end{equation}
for $\rho^{(G)}>>p^{(G)}$ with $\Lambda^{(b,G)}=c_2^{(b)}+c_2^{(G)}$. Here, we set $\Lambda^{(b,G)}=0$ as a cosmological constant.

\subsection{Model I}
Using Eqs. \eqref{13} and \eqref{17} for the functional form \eqref{20}, the expression for the Hubble parameter read as

\begin{multline}
\label{22}
H(a)=H_0 \left[\frac{\Omega_0^{(b)} \left(a^{3/2} m-1\right)}{a^3}-\Omega_0^{(G)} \left((1-k) a^{-3 (\alpha +1)}+k\right)^{\frac{1}{\alpha +1}}\right.\\
 \left.\left(1- \sqrt{12}\, k^{3/2}\,n_1 \left(\left((1-k) a^{-3 (\alpha +1)}+k\right)^{\frac{1}{\alpha +1}}\right)^{-\frac{3\alpha+2}{2}}\right)\right]^{\frac{1}{2}},
\end{multline} 
where $m=\frac{c_1^{(b)}\rho_0^{-\frac{1}{2}}}{8\pi G}$ and $n_1=\frac{c_1^{(G)}\rho_0^{\frac{1}{2}}}{8\pi G}$.\\
Note that the dimensionless present density parameter $\Omega_0^{(b)}$ and $\Omega_0^{(G)}$ are dependent on each other. Applying the present value $H(a_0)=H_0$ ($a_0=1$) we get \cite{Shabani/2017},
\begin{equation}
\label{23}
\Omega_0^{(G)}=\frac{\Omega_0^{(b)}(m-1)-1}{1-\sqrt{12}\,n_1\,k^{\frac{3}{2}}}.
\end{equation}
Furthermore, the relation $a=\frac{1}{1+z}$, gives the Hubble parameter versus redshift as
\begin{widetext}
\begin{multline}
\label{24}
H(z)=H_0 \left[\Omega_0^{(b)} (z+1)^3 \left(m \left(\frac{1}{z+1}\right)^{3/2}-1\right)-\Omega_0^{(G)} \left((1-k) \left(z+1\right)^{3 (\alpha +1)}+k\right)^{\frac{1}{\alpha +1}}\right.\\\left.
\left(1-2 \sqrt{3}\,k^{3/2}\,n_1 \left(\left((1-k) \left(z+1\right)^{3 (\alpha +1)}+k\right)^{\frac{1}{\alpha +1}}\right)^{-\frac{3\alpha+2}{2}}\right)\right]^{\frac{1}{2}}.
\end{multline}
\end{widetext}

\subsection{Model II}
Similarly, Eqs. \eqref{13} and \eqref{17} with Eq.\eqref{21}, gives the Hubble parameter as
\begin{widetext}
\begin{multline}
\label{25}
H(a)=H_0 \left[\frac{\Omega_0^{(b)} \left(a^{3/2} m-1\right)}{a^3}-\Omega_0^{(G)} \left((1-k) a^{-3 (\alpha +1)}+k\right)^{\frac{1}{\alpha +1}} 
\left(1- 2\,n_2 \left(\left((1-k) a^{-3 (\alpha +1)}+k\right)^{\frac{1}{\alpha +1}}\right)^{-\frac{1}{2}}\right)\right]^{\frac{1}{2}}.
\end{multline}
\end{widetext}
where $m=\frac{c_1^{(b)}\rho_0^{-\frac{1}{2}}}{8\pi G}$ and $n_2=\frac{c_1^{(G)}\rho_0^{-\frac{1}{2}}}{8\pi G}$.\\
Note that the dimensionless present density parameter $\Omega_0^{(b)}$ and $\Omega_0^{(G)}$ are dependent on each other. Applying the present value $H(a_0)=H_0$ ($a_0=1$), we have \cite{Shabani/2017}
\begin{equation}
\label{26}
\Omega_0^{(G)}=\frac{\Omega_0^{(b)}(m-1)-1}{1-2\,n_2}.
\end{equation}
The expression for the Hubble parameter in terms of redshift reads
\begin{widetext}
\begin{multline}
\label{27}
H(z)=H_0 \left[\Omega_0^{(b)} (z+1)^3 \left(m \left(\frac{1}{z+1}\right)^{3/2}-1\right)-\Omega_0^{(G)} \left((1-k) \left(z+1\right)^{3 (\alpha +1)}+k\right)^{\frac{1}{\alpha +1}}\right.\\\left.
\left(1-2\,n_2 \left(\left((1-k) \left(z+1\right)^{3 (\alpha +1)}+k\right)^{\frac{1}{\alpha +1}}\right)^{-\frac{1}{2}}\right)\right]^{\frac{1}{2}}.
\end{multline}
\end{widetext}

\section{observational data}
\label{section 3}
This section addresses the constraints on Model I and Model II. Three types of observational data will be used to break the degeneracy between the model parameters: the Hubble data (OHD) \cite{Yu/2018,Farooq/2017,Moresco/2015}, Type Ia Supernovae (SNeIa) \cite{Scolnic/2018}, and baryon acoustic oscillation (BAO) \cite{Blake/2011,Percival/2010}.\\
For SNeIa data, we use the Pantheon sample, which consists of 1048 data points from the Pan-STARSS1 (PS1) Medium Deep Survey, the Low-z, SDSS, SNLS, and HST surveys \cite{Chang/2019}. \\
We use the total likelihood function to get the joint constraints on model parameters from the aforementioned cosmic probes.

\begin{equation*}
\mathcal{L}_{tot}=  \mathcal{L}_{OHD} \times \mathcal{L}_{SNe Ia} \times \mathcal{L}_{BAO}.
\end{equation*}

Henceforth, the relevant $\chi^{2}_{tot}$ is given as 

\begin{equation}
\label{28}
\chi^{2}_{tot} = \chi^{2}_{OHD} + \chi^{2}_{SNe Ia} + \chi^{2}_{BAO}. 
\end{equation}

Finally, we use the Markov Chain Monte Carlo (MCMC) sample to explore the parameter space from the python package $emcee$ \cite{Mackey/2013} for likelihood minimization, which is widely used in astrophysics and cosmology.

\subsection{OHD}

We use a standardized compilation containing 31 measurements obtained using the differential age approach. One can measure the rate of expansion of the cosmos at redshifts $z$ using the differential age method. The chi-square ($\chi^{2}$) for $OHD$ is calculated as follows:

\begin{equation}
\label{29}
\chi^{2}_{OHD}= \sum_{i=1}^{31} \frac{\left[ H(z_{i}, \theta) - H_{obs} (z_{i})\right]^{2} }{\sigma^{2}(z_{i})},
\end{equation}
where $H_{obs} (z_{i})$ is the observational value, $\sigma(z_{i})$ stand for the observational error, $H(z_{i})$ is the theoretical value for a given model at redshifts $z_{i}$, and $\theta$ is the parameter space.

\subsection{SNeIa}

The standard candles and, most likely, the $SNe Ia$ are the common types of cosmological probes. The present analysis uses the recent $SNe Ia$ dataset, the Pantheon sample, which contains 1048 points of distance moduli $\mu^{obs}$ at various redshifts $z_{i}$ in the range $0.01 < z_{i} < 2.26$. The corresponding $\chi^{2}$ is defined as 
\begin{equation}
\label{30}
\chi^{2}_{SN}= min \sum_{i,j=1}^{1048} \Delta \mu_{i} (C^{-1}_{SN})_{ij} \Delta \mu_{j}.
\end{equation}
Here, $\Delta \mu_{i}= \mu^{th}(z_{i},\theta)-\mu_{i}^{obs}$, where $\theta$ is a parameter space, $C_{SN}$ is the covariance matrix.\\
Considering the Tripp estimator \cite{Tripp/1998}, the standardized observational $SNe Ia$ distance modulus is given by 
\begin{equation}
\label{31}
\mu = m_{B} - M_{B} + \alpha x_{1} - \beta c + \Delta_M + \Delta_B,
\end{equation} 
where $m_{B}$ refers to the observed  peak magnitude at the time of B-band maximum, $M_{B}$ is the absolute magnitude, $c$ describes the $SNe Ia$ color at maximum brightness. The parameters $\alpha$ and $\beta$ corresponds to the relation between luminosity-stretch and luminosity-colour, respectively, and $x_{1}$ is a light curve shape parameter. Furthermore, $\Delta_M$ and $\Delta_B$ are distance corrections based on the host galaxy's mass and simulation-based expected biases, respectively.\\
According to the new approach called BEAMS with Bias Correction (BBC) \cite{Kessler/2017}, the nuisance parameters in the Tripp formula were retrieved, and the observed distance modulus is reduced to the difference between the corrected apparent magnitude $M_{B}$ and the absolute magnitude $m_{B}$, i.e. $\mu = m_{B} - M_{B}$. 
Henceforth, the expression for $\mu^{th}(z)$ is expressed as 
\begin{equation}
\label{32}
\mu^{th}(z)= 5 log_{10}\left[ \frac{d_{L}(z)}{1 Mpc}\right] +25, 
\end{equation}
where 
\begin{equation}
\label{33}
d_{L}(z)= c(1+z) \int_{0}^{z} \frac{dx}{H(x,\theta)}.
\end{equation}

\subsection{BAO}
For BAO data, we adopt the compilation from $6dFGS$, $SDSS$ and $Wiggle~~Z$ surveys at different redshifts. To obtain BAO constraints, we use $\frac{d_{A}(z_*)}{D_{v}(z)}$ and the following cosmology
\begin{eqnarray}
\label{34}
d_{A}(z) &=& c \int_{0}^{z} \frac{dz^{'}}{H(z^{'})}\\
D_{v}(z) &=& \left[ \frac{d_{A}^{2}(z) c z}{H(z)}\right] ^{1/3}
\end{eqnarray}
where $z_*$ is the photon decoupling redshift, $d_{A}(z)$ is the comoving angular diameter distance, and $D_{v}$ is the dilation scale.
The $\chi^{2}$ for BAO is defined as  
\begin{equation}
\label{36}
\chi^{2}_{BAO}= X^{T} C^{-1}_{BAO} X,
\end{equation}
where $X$ depends on the survey considered and $C$ is the covariance matrix \cite{Giostri/2012}.
 
\subsection{Observational Results}

The constraints on model parameters of Model I and Model II from the combined $OHD+SNeIa+BAO$ are given by minimizing $\chi^{2}_{OHD}+\chi^{2}_{SNeIa}+\chi^{2}_{BAO}$. The results are obtained as in Table I. Further, $1-\sigma$ and $2-\sigma$ likelihood contours for Model I and Model II for the possible subsets of parameter space $(k,m,n_{1}/n_{2},\alpha)$ is presented in Figs. \ref{figure 1} and \ref{figure 2}, respectively.

\begin{widetext}

\begin{figure}[H]
\centering
\includegraphics[scale=0.7]{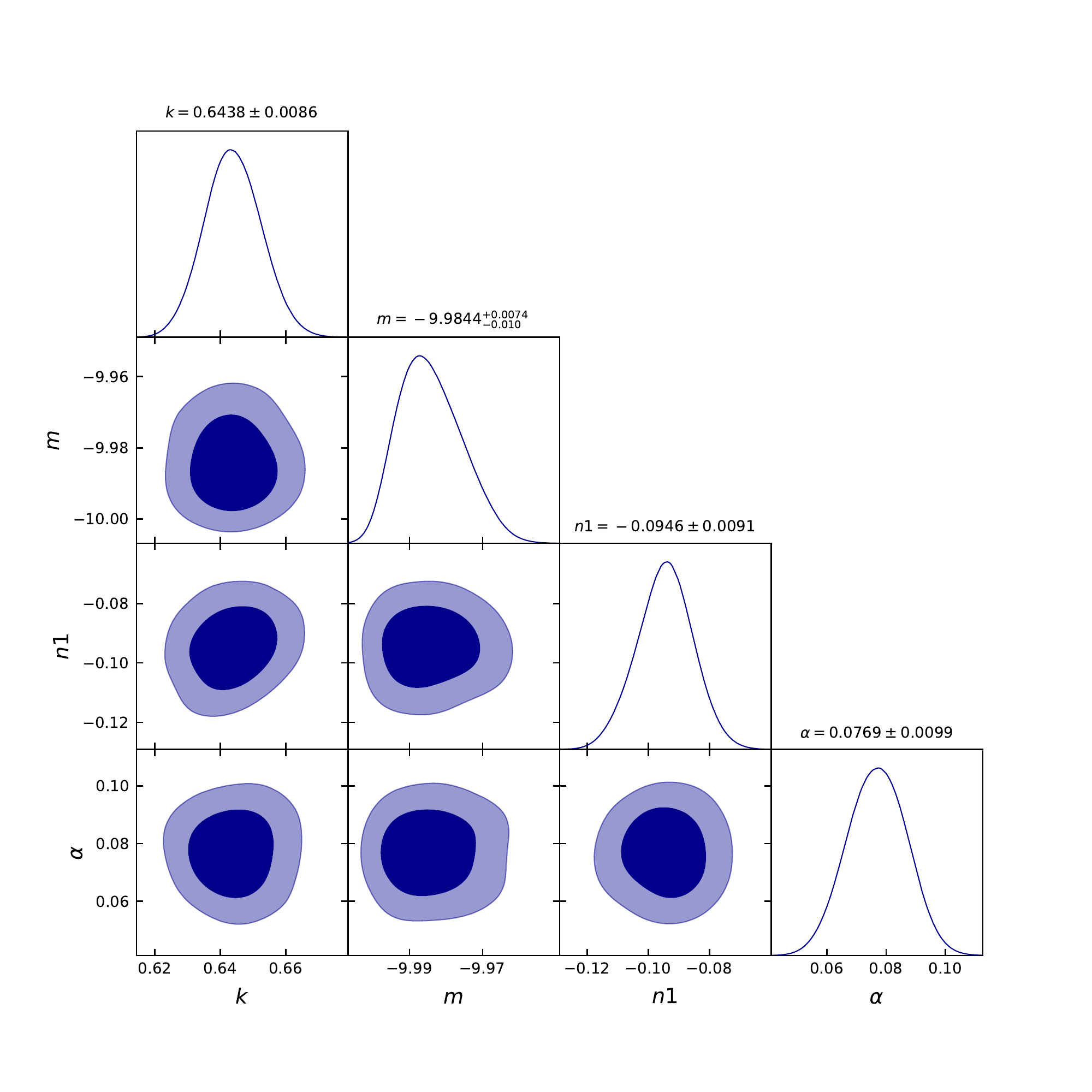}
\caption{The constraints on Model I from $OHD+SNeIa+BAO$ corresponding to $1-\sigma$ and $2-\sigma$ confidence regions.}
\label{figure 1}
\end{figure}
\begin{figure}[H]
\centering
\includegraphics[scale=0.65]{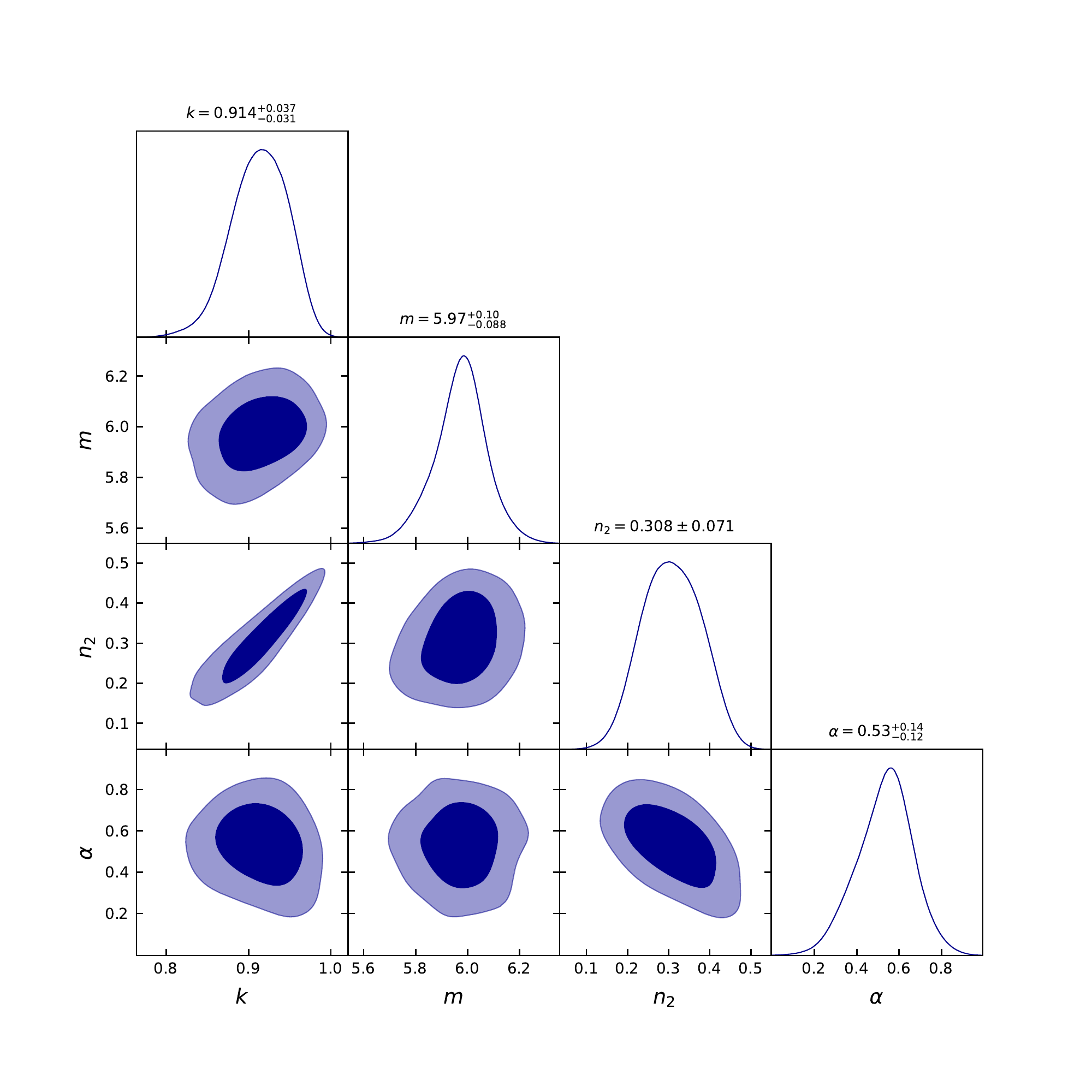}
\caption{The constraints on Model II from $OHD+SNeIa+BAO$ corresponding to $1-\sigma$ and $2-\sigma$ confidence regions.}
\label{figure 2}
\end{figure}

\begin{table}[h!]
\begin{center}
  \caption{Table showing the marginalized constraints on model parameters for Model I and Model II of combined $OHD+SNeIa+BAO$.}
    \label{table1}

\begin{tabular}{|l|c|c|c|c|}
\hline 
Models              & k & m & $n_{1}$ and $n_{2}$ & $\alpha$ \\
\hline
Model I             & $0.6438^{+0.0086}_{-0.0086}$ & $-9.9844^{+0.0074}_{-0.0100}$ & $n_{1} = -0.0946^{+0.0091}_{-0.0091}$ & $0.0769^{+0.0099}_{-0.0099}$ \vspace{0.1cm} \\
\vspace{0.1cm} 
Model II           & $0.914^{+0.037}_{-0.031}$ & $5.97^{+0.010}_{-0.088}$ & $n_{2}=0.308^{+0.071}_{-0.071}$ & $0.53^{+0.14}_{-0.12}$   \\
\hline
\end{tabular}
\end{center}
\end{table}

\end{widetext}

\section{cosmological parameters}
\label{section 4}

\begin{widetext}

\begin{figure}[H]
\centering
\begin{subfigure}{0.45\textwidth}
    \includegraphics[width=\textwidth]{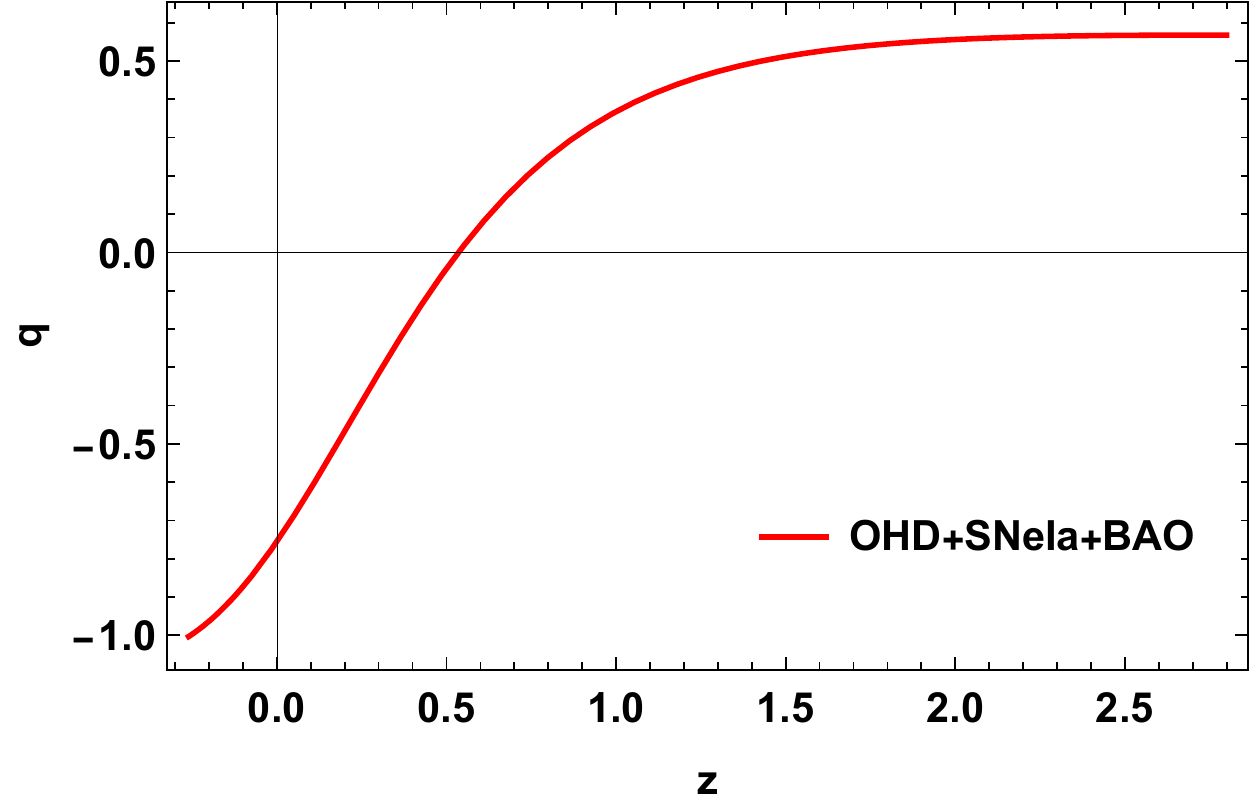}
    \caption{Plot of $q$ versus $z$ for model I}
    \label{figure 3}
\end{subfigure}
\hfill
\begin{subfigure}{0.45\textwidth}
    \includegraphics[width=\textwidth]{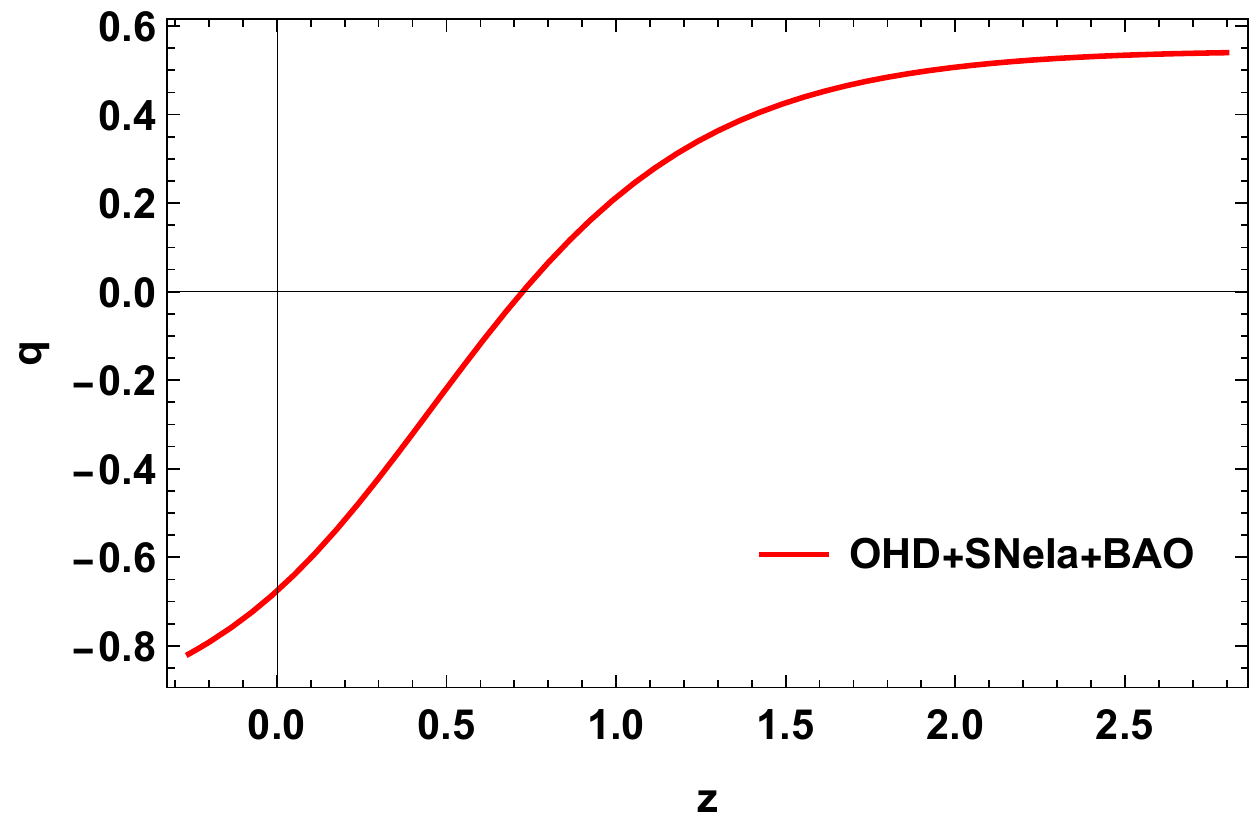}
    \caption{Plot of $q$ versus $z$ for model II}
    \label{figure 4}
\end{subfigure}
\hfill
\begin{subfigure}{0.45\textwidth}
    \includegraphics[width=\textwidth]{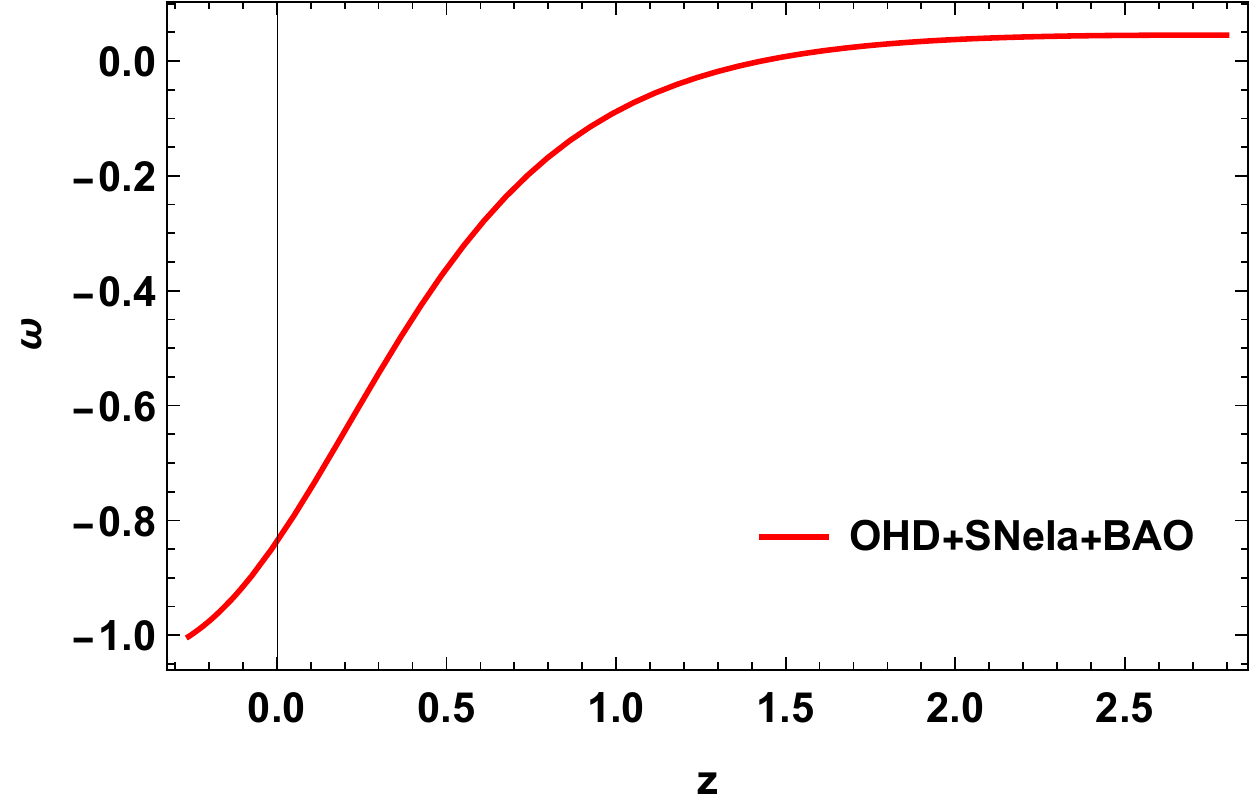}
    \caption{Plot of $\omega$ versus $z$ for model I}
    \label{figure 5}
\end{subfigure}
\hfill
\begin{subfigure}{0.45\textwidth}
    \includegraphics[width=\textwidth]{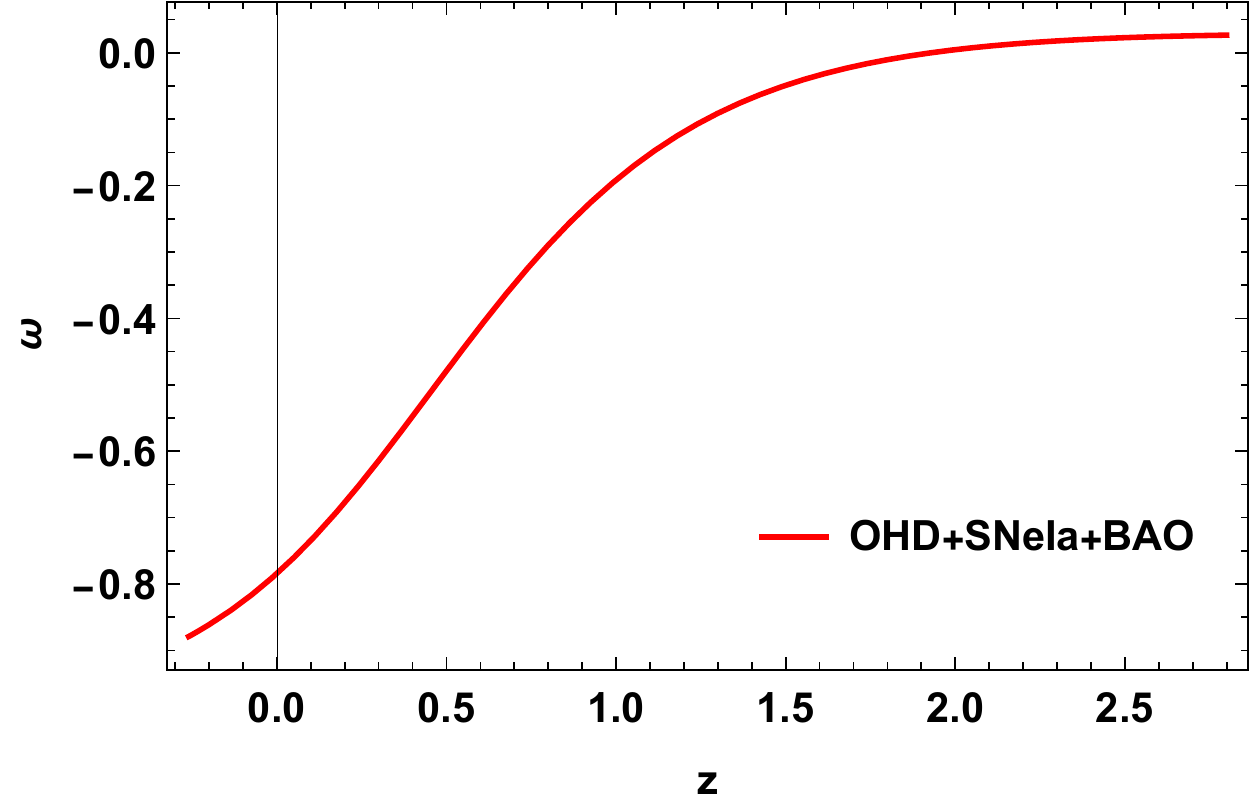}
    \caption{Plot of $\omega$ versus $z$ for model II}
    \label{figure 6}
\end{subfigure}
\caption{Evolution of $q$ and $\omega$ as a function of redshift $z$ using the constrained values of model parameters for Model I and Model II.}
\label{Parameters}
\end{figure}

\end{widetext}

The deceleration parameter $q$ is a crucial cosmological variable that indicates whether the evolution of the universe is accelerating or decelerating. A positive value of $q$ indicates decelerated expansion, while a negative $q$ indicates an accelerated expansion. The deceleration parameter $q$ in terms of the Hubble parameter $H$ is defined as $q=-\frac{\dot{H}}{H^2}-1$
The deceleration parameter shown in Figs. \ref{figure 3} and \ref{figure 4} starts with positive values as a function of redshift $z$ for Model I and Model II, respectively. However, acceleration arises at late-times, i.e. $q$ crossing $0$ at $z_{t}$.
The results obtained are as follows: we find the value of the transition redshift as $z_t=0.53^{+0.004}_{-0.003}$ \cite{Jesus/2020,Garza/2019}, and the present value of the deceleration parameter is $q_0=-0.75^{+0.01}_{-0.01}$ \cite{Virey/2005,Lua/2009,Garza1/2019} for Model I corresponding to the model parameters constrained by the $OHD+SNeIa+BAO$. In case of Model II, the values are obtained as  $z_t=0.71^{+0.11}_{-0.03}$ \cite{Farooq/2013,Santos/2016,Farooq/2017,Mamon/2017,Mamon/2018}, and $q_0=-0.67^{+0.01}_{-0.02}$ \cite{Lua/2009,Segio/2012,Arora/2022} for $OHD+SNeIa+BAO$ . Henceforth, the Chaplygin gas model appears to yield a deceleration parameter, which describes the transition phases of the universe from deceleration to acceleration.

It is suitable to define an effective EoS in the higher-order gravity to explain the current accelerated expansion of the universe through the consequences of the geometrically modified gravity terms in the Friedmann equations. The EoS parameter can be useful in investigating the nature of the component that dominates the universe. The EoS parameter is defined as $\omega=\frac{p}{\rho}$, where $p$ is the pressure and $\rho$ is the energy density.\\
We can find the evolution of $\omega$ in  Figs. \ref{figure 5} and \ref{figure 6} for Model I and Model II, respectively. Fitting the model to observation data provides us the present value of $\omega$ as $w_0=-0.84^{+0.006}_{-0.010}$ for model I and $w_0=-0.78^{+0.01}_{-0.02}$ for model II \cite{Arora/2022,Mandal/2021}. 
It is seen that $\omega<0$ and lies in the quintessence region showing the present accelerating phase. 
Finally, the parameter indicates that the dark energy sector behaves like a quintessence at present and tends to -1 (i.e., cosmological constant) in the near future.

\section{Statefinder parameters} \label{section 5}
\begin{figure}[H]
\includegraphics[scale=0.45]{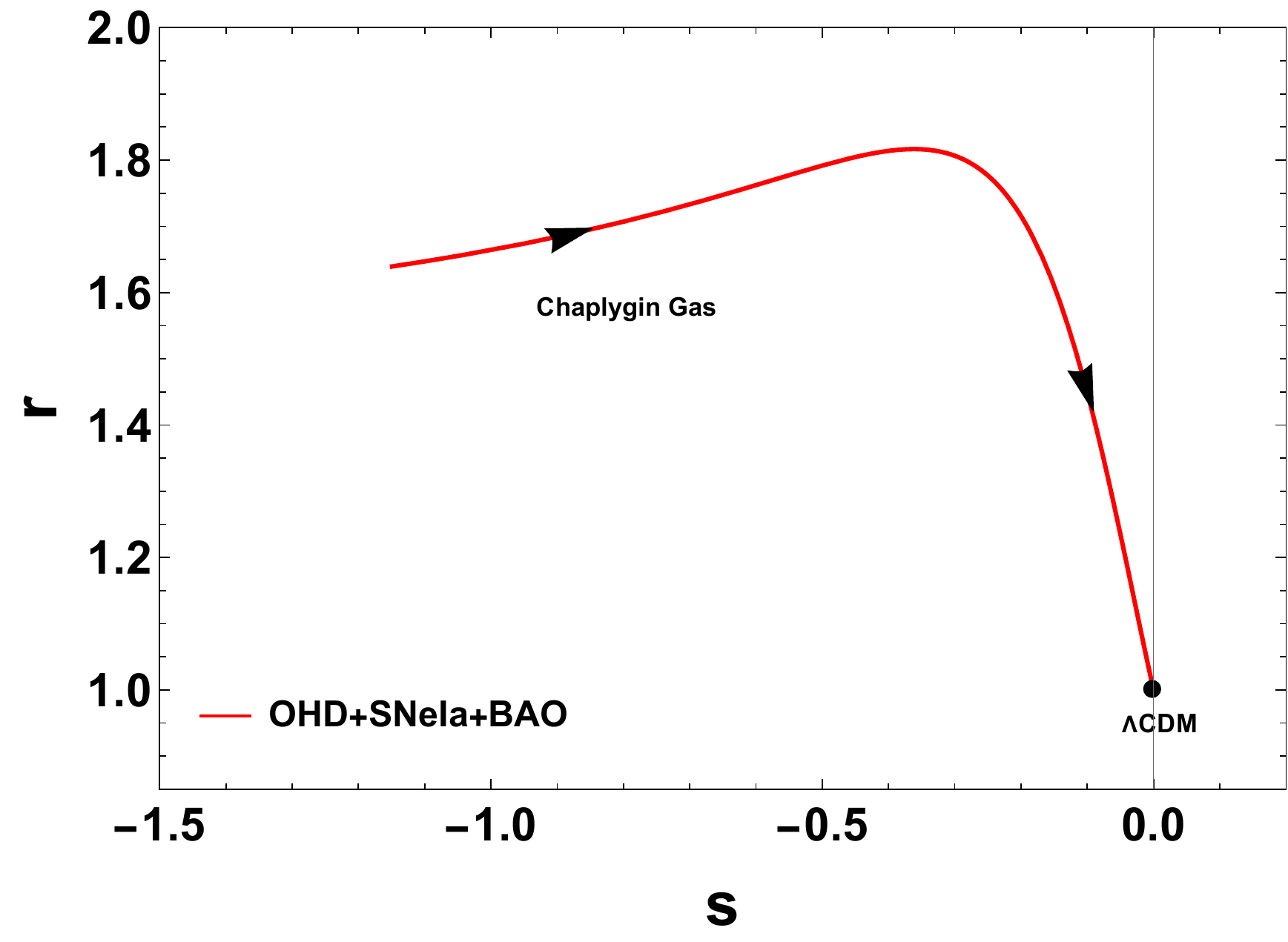}
\caption{Trajectory of  $s-r$ plane for the model I.}
\label{figure 7}
\end{figure}

\begin{figure}[H]
\includegraphics[scale=0.45]{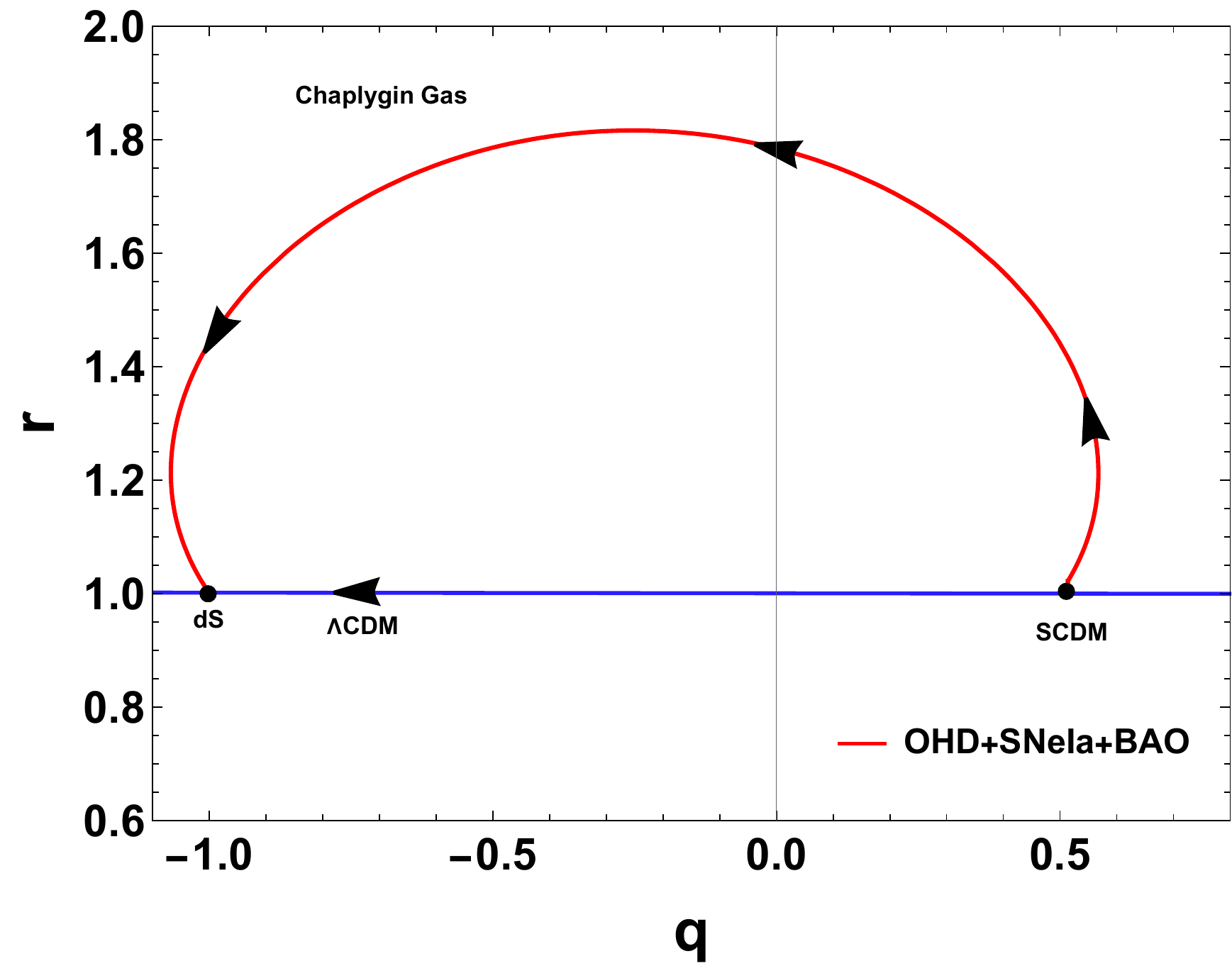}
\caption{Trajectory of  $q-r$ plane for the model I.}
\label{figure 8}
\end{figure}

\begin{figure}[H]
\includegraphics[scale=0.44]{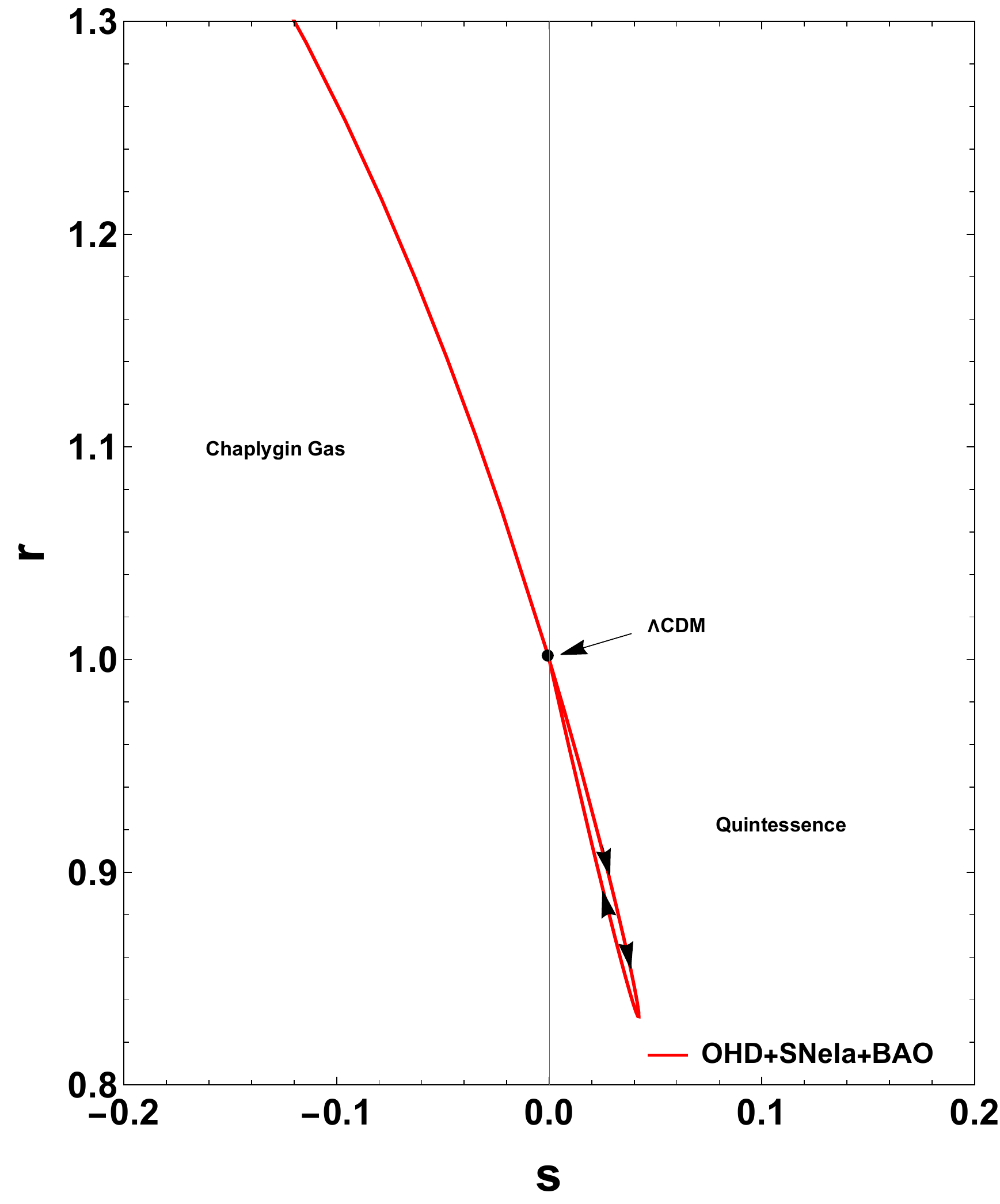}
\caption{Trajectory of  $s-r$ plane for the model II.}
\label{figure 9}
\end{figure}

\begin{figure}[H]
\includegraphics[scale=0.45]{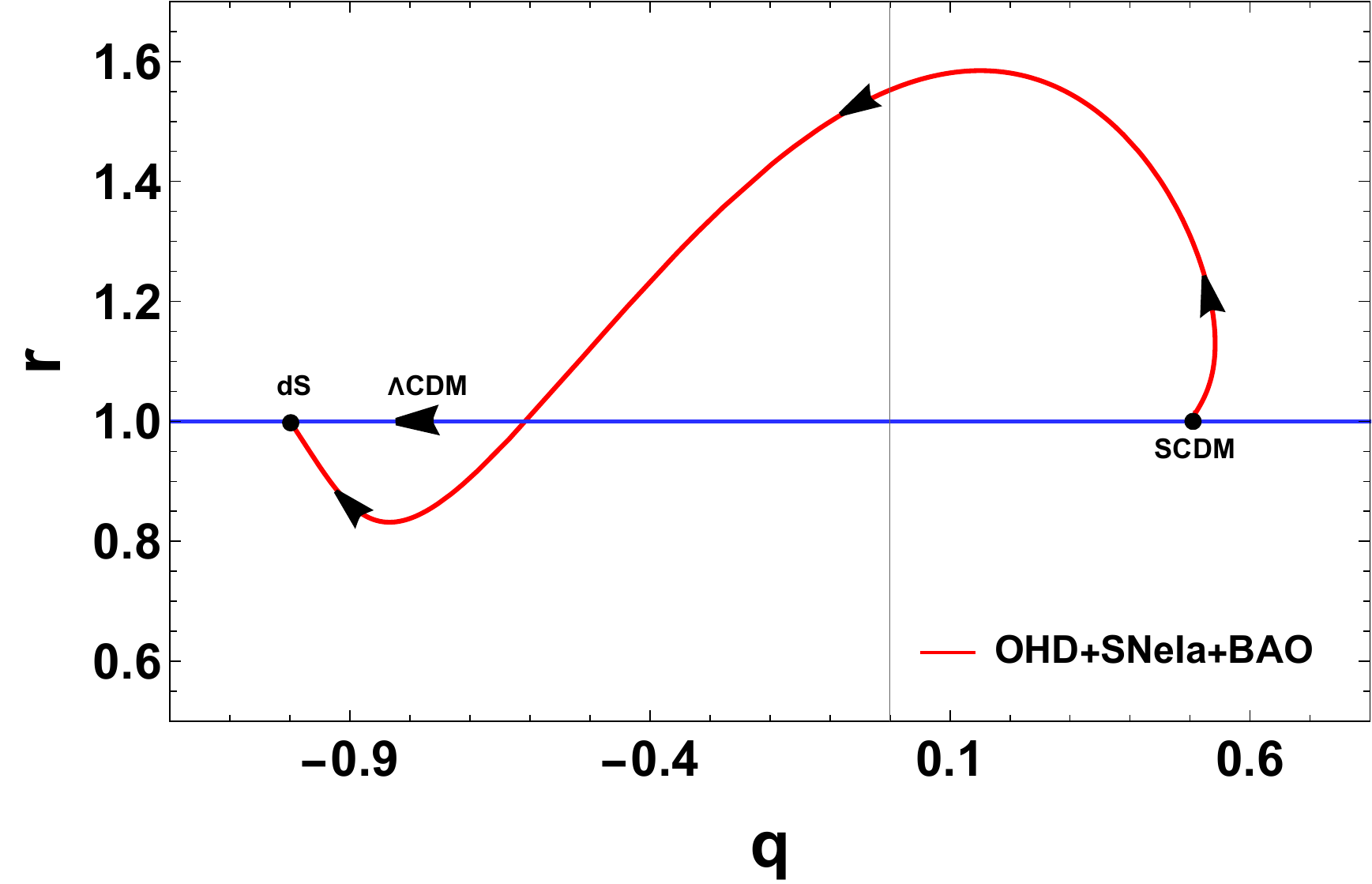}
\caption{Trajectory of  $q-r$ plane for the model II.}
\label{figure 10}
\end{figure}

The challenge of discriminating amongst the various DE candidates has arisen as a result of the development of numerous DE models to comprehend cosmic acceleration. For this purpose, Sahni et al. \cite{Sahni/2003,Alam/2003} introduced a geometrical diagnostic called the statefinder diagnostic, which is used to characterize the properties of the DE. Mathematically, the statefinder pair is defined as 
\begin{equation}
\label{37}
r=\frac{\dddot{a}}{aH^3}
\end{equation}
\begin{equation}
\label{38
}
s=\frac{r-1}{3(q-\frac{1}{2})}
\end{equation}
It is determined by the second and third-order derivatives of the expansion factor. Here, we perform the statefinder diagnostic for two $f(Q,T)$ models. In Figs. \ref{figure 7} and \ref{figure 8}, we show the diagnostic results for the constrained values of parameters from $OHD+SNeIa+BAO$ for Model I. Similarly, Figs. \ref{figure 9} and \ref{figure 10} shows the evolution trajectory of $s-r$ and $q-r$ pairs, respectively for Model II.\\
In Fig. \ref{figure 7}, the trajectory of $s-r$ pair always lies in Chaplygin gas regime $(r>1, s<0 )$  and converges to the $\Lambda$CDM point $(0,1)$ in future. Further, the trajectory of $q-r$ plane in Fig. \ref{figure 8} starts at $SCDM$ $(0.5,1)$ and approaches the de sitter $(dS)$ point $(-1,1)$ at late times,  with the Chaplygin gas behaviour throughout the evolution.
The evolution of $s-r$ plane for model II is shown in Fig. \ref{figure 9} which originates in the Chaplygin gas regime and enters the quintessence zone ($r<1,s>0$) before attaining the $\Lambda$CDM point. 
On the other, trajectory in the $q-r$ plane starts at the $SCDM$ point and enters the quintessence zone before approaching the de Sitter point, as shown in Fig. \ref{figure 10}. The present value of $(s,r)$ parameter is $(-0.18,1.68)$ for model I and $(0.02,0.92)$ for model II, corresponding to the values of model parameters constrained by $OHD+SNeIa+BAO$ \cite{Gorini/2003,Mamon/2021,Gadbail/2021}.
  

\section{Conclusion}
\label{section 6}
Within the framework of modified gravity, we adopted a unique geometrical scenario based on the non-metricity and the trace of the energy-momentum tensor, i.e., the $f(Q,T)$ gravity. In this article, we have investigated the cosmological implications of the $f(Q, T)$ gravity theory by imposing the  GCG coupled with a pressureless baryonic matter in the trace of the energy-momentum tensor. 

We consider the $f(Q,T)$ function as $f(Q,T)=Q+h(T^{(b,G)})$, which is linear in $Q$ and non-linear in $T$. Here, $h(T^{(b,G)})$ is an unknown function of $T$, which is determined using the field equations of $f(Q,T)$ and the Bianchi identity. Further, we obtained two different functional forms of $f(Q,T)$ model in two separated cases for GCG.\\
The GCG model I in the high pressure regimes, i.e. $p^{(G)}>>\rho^{(G)}$ and the model II of GCG in the high density regimes, i.e. $\rho^{(G)}>>p^{(G)}$. After obtaining and normalizing the Hubble parameter for each model, we constrained the model parameters using the recent observational data $OHD+SNeIa+BAO$. We assumed the present density of the baryonic matter as $\Omega_0^{(b)}=0.05$ \cite{Mamon/2021}.\\
We started testing our cosmological solutions in section \ref{section 3} and have obtained the best fit values of model parameters as listed in Table \ref{table1}. Corresponding to the constrained values, the deceleration parameter $q$ shows that the universe experiences a transition from decelerated to accelerating phases of the universe at a redshift value $z_t=0.53^{+0.004}_{-0.003}$ and $z_t=0.71^{+0.11}_{-0.03}$ with the present values $q_0=-0.75^{+0.01}_{-0.01}$ and $q_0=-0.67^{+0.01}_{-0.02}$ for Model I and Model II, respectively. 
Furthermore, the EoS parameter presented shows that the universe behaves like a quintessence dark energy. The current value of the EoS parameter is $w_0=-0.84^{+0.006}_{-0.010}$ and $w_0=-0.78^{+0.01}_{-0.02}$ for Model I and Model II, respectively. \\
Finally, in section \ref{section 5}, the evolutionary trajectory of the statefinder pairs $s-r$ and $q-r$ are wholly lying in the Chaplygin gas regime and approaches to the $\Lambda$CDM point at late-times for Model I. We can observe that the statefinder diagnostic is a better quantity than EoS to distinguish DE models in this case. In view of Model II, the statefinder pairs $s-r$ and $q-r$ evolutionary trajectory comes from the Chaplygin gas regime at the early times and enters the quintessence region before approaching the $\Lambda$CDM point. \\
It is seen that both the models are useful in differentiating different dark energy models, but the results from observational constraints on all the cosmological parameters such as deceleration parameter, equation of state, and statefinder pairs are in favor of Model I.  The present model is therefore, a good alternative to study the dynamics of the universe.

\section*{Data Availability Statement}

There are no new data associated with this article.

\section*{Acknowledgments}

GNG acknowledges University Grants Commission (UGC), New Delhi, India for awarding Junior Research Fellowship (UGC-Ref. No.: 201610122060). SA acknowledges CSIR, Govt. of India, New Delhi, for awarding Senior Research Fellowship. PKS acknowledges CSIR, New Delhi, India for financial support to carry out the Research project [No.03(1454)/19/EMR-II Dt.02/08/2019]. We are very much grateful to the honorable referee and the editor for the illuminating suggestions that have significantly improved our work in terms of research quality and presentation.

\end{document}